\documentclass[a4paper,12pt]{article}
\pdfoutput=1
\pdfmapfile{=cm.map}
\pdfmapfile{+symbols.map}
\pdfmapfile{+cmextra.map}
\usepackage{amsmath,amsfonts,amssymb,multicol,slashed}
\usepackage{bm,booktabs,graphicx}
\usepackage[dvipsnames]{xcolor}
\usepackage[colorlinks=true,linkcolor=blue!50!black,
            urlcolor=red!60!black,citecolor=green!40!black]{hyperref}

\newcommand{\keV}{\,\mathrm{keV}}
\newcommand{\MeV}{\,\mathrm{MeV}}
\newcommand{\GeV}{\,\mathrm{GeV}}
\newcommand{\TeV}{\,\mathrm{TeV}}
\newcommand{\fig}[1]{~\ref{fig:#1}}

\makeatletter

\font\ital=cmu10

\def\hhref#1{\href{http://arxiv.org/abs/#1}{arXiv:#1}}
\usepackage{xstring}
\newcommand{\hhrefq}[1]{\IfSubStr{#1}{:}{\href{http://inspirehep.net/search?ln=en&ln=en&p=#1&of=hb&action_search=Search&sf=&so=d&rm=&rg=25&sc=0}{InSpire:#1}}{\hhref{#1}}}

\def\art{\@ifnextchar[{\eart}{\oart}}
\def\eart[#1]#2#3#4#5#6{{\rm #2}, {\em #3 \bf #4} {\rm (#6) #5} ({\em #1})}
\def\article{\@ifnextchar[{\earticle}{\oarticle}}
\def\oarticle#1#2#3#4#5#6{{\rm #1}, {\ital `#6'}, {\rm #2 #3 (#5) #4}}
\def\earticle[#1]#2#3#4#5#6#7{{\rm #2}, {\ital `#7'}, {\rm #3 #4 (#6) #5}  [\hhrefq{#1}]}
\def\hepart[#1]#2{{\rm #2, \sl#1}}
\def\heparticle[#1]#2#3{#2, {\ital `#3'} [\hhrefq{#1}]}
\newcommand{\hhrefqq}[1]{\IfBeginWith{#1}{10.}{\href{https://doi.org/#1}{doi:#1}}{\hhrefq{#1}}}
\def\earticle[#1]#2#3#4#5#6#7{{\rm #2}, {\ital `#7'}, {\rm #3 #4 (#6) #5}  [\hhrefqq{#1}]}
\renewenvironment{thebibliography}[1]
{\begin{multicols}{2}[\section*{\refname}]%
		\@mkboth{\MakeUppercase\refname}{\MakeUppercase\refname}%
		\list{\@biblabel{\@arabic\c@enumiv}}%
		{\settowidth\labelwidth{\@biblabel{#1}}%
			\leftmargin\labelwidth
			\advance\leftmargin\labelsep
			\@openbib@code
			\usecounter{enumiv}%
			\let\p@enumiv\@empty
			\renewcommand\theenumiv{\@arabic\c@enumiv}}%
		\sloppy
		\clubpenalty4000
		\@clubpenalty \clubpenalty
		\widowpenalty4000%
		\sfcode`\.\@m}
	{\def\@noitemerr
		{\@latex@warning{Empty `thebibliography' environment}}%
		\endlist\end{multicols}}

\newcommand{\beq}{\begin{equation}}
\newcommand{\eeq}{\end{equation}}

\begin{document}
\begin{flushright}
OU-HET-1325
\end{flushright}
\begin{center}
\bigskip
{\bf\LARGE\color{red!60!black} Neutron disappearance and \\[0.2ex]   the LZ nuclear recoil event}\\[3ex]
{\bf Mohammad Aghaie}$^a$ and {\bf Alessandro Strumia}$^b$\\[2ex]
{\it $^a$ Department of Physics, The University of Osaka, Japan\\
$^b$ Dipartimento di Fisica, Universit\`a di Pisa, Italia}\\[2ex]
{\bf\large\color{blue!20!black} Abstract}
\begin{quote}\large\color{blue!20!black} 
The LZ experiment reported a nuclear recoil with anomalous energy $E_R=248\keV$,
around the typical neutron motion in xenon.
This motivates interpreting the event as the disappearance of a bound neutron, 
either through spontaneous decay or a dark-matter-induced reaction.
If the energy released is large, the predicted recoil spectrum is broad around the observed value;
only $\sim10\%$ of xenon events avoid extra nuclear activity;
the disappearance rate is target-independent and mildly constrained by oxygen experiments.
In the special case of near-threshold DM-induced neutron disappearance,
it can be open in xenon and closed in oxygen,
the quiet fraction rises to $\sim30\%$,  the recoil spectrum turns into lines.
\end{quote}
\end{center}
\thispagestyle{empty}

\bigskip

\noindent
The LZ experiment found in an exposure of $2.84$ ton yr~\cite{2609.02823}
one  nuclear-recoil candidate with energy  $ E_R=(248\pm33)\keV$
well above the recoil scale expected from ordinary halo-DM scattering.
This anomalously high energy implies a global significance of $2.6\sigma$ and motivates considering different sources.
The corresponding nuclear momentum $p \approx \sqrt{2M_A E_R}\approx 245\MeV$
is the typical momentum of one neutron in a nucleus.
Indeed, in the simplest uniform Fermi-gas approximation, nuclear neutron momenta are filled up to the Fermi sphere
$p_{\rm Fermi} \approx (9\pi N/4A)^{1/3}/r_0 \approx 270 \MeV$  where $r_0 \approx 1.2\,{\rm fm}$ is the nuclear radius parameter
and $A=N+Z$ with $Z=54$,  $N \approx 77.3$  for xenon.

\smallskip

\begin{figure}[t]
$$\includegraphics[width=0.47\textwidth]{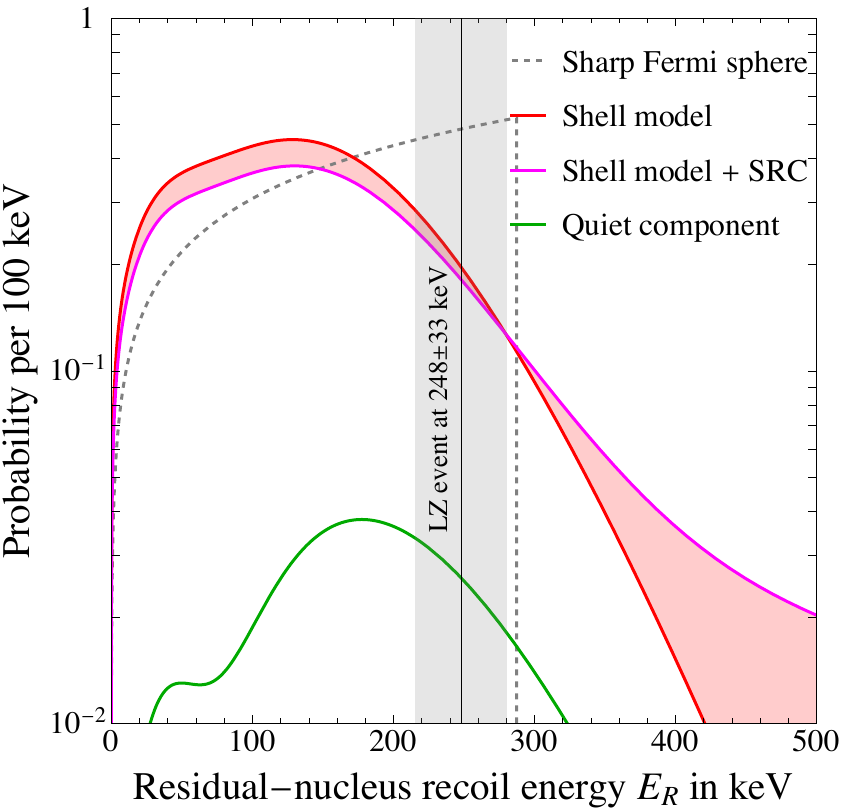}\quad
\includegraphics[width=0.47\textwidth]{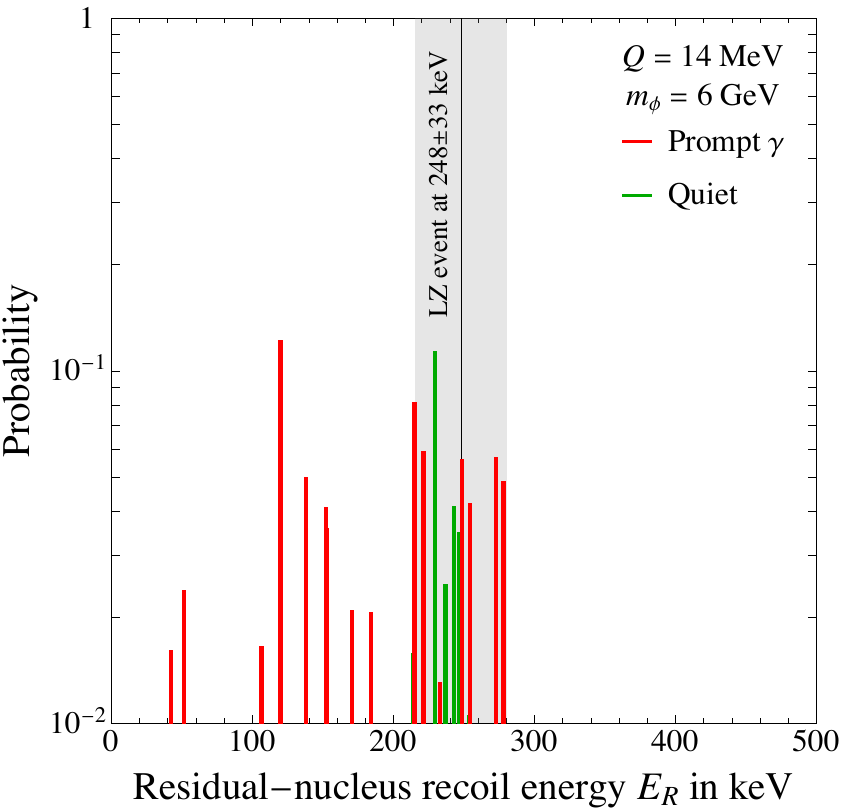}$$
\begin{center}
\caption{\em\label{fig:LZn} Predicted nuclear recoil energy spectrum from invisible neutron decay in xenon.
Left: assuming large released energy, and using:
a naive sharp Fermi sphere with $p_{\rm Fermi}=265\MeV$ (dashed); a shell model computation,
possibly improved by including Short Range Correlations that add a $1/p^4$ tail.
The uncertain green curve shows our estimate for the fraction of events with no prompt nuclear activity.
Right: near threshold induced neutron disappearance. 
}
\end{center}
\end{figure}

This coincidence invites trying to interpret the LZ event as a process in which one bound neutron disappears into invisible particles.
In the impulse approximation the remaining nucleus is a spectator, so this interpretation
explains the observed recoil energy.\footnote{A different process, incoherent dark-matter absorption
$\chi n\to\nu n$, ejects an energetic neutron and was shown to be
constrained by KamLAND~\cite{2609.01592}.
Other somehow related LZ interpretations include 
inelastic dark matter~\cite{2609.01475},
exothermic and endothermic dark matter~\cite{2609.04673}, 
strongly modulated near-threshold scattering~\cite{2609.04181}, 
atmospheric-neutrino upscattering~\cite{2609.04185},  
momentum-suppressed elastic scattering through a pseudoscalar portal~\cite{2609.04186}.}
Neutron disappearance can happen either because the neutron decays
(the possible invisible neutron decay channels $n\to \slashed{E}$ have been classified in~\cite{2510.14940})
or because the neutron is hit by a Dark Matter particle 
(possible scattering channels ${\rm DM}\, n \to \slashed{E}$ have been considered e.g.\ in~\cite{2112.09111,2406.00445,2511.18722}, in mass ranges not adopted here).
The left panel of fig.\fig{LZn} shows the predicted recoil energy spectrum assuming invisible neutron decay.
Taking into account that shell structure and nuclear correlations smooth the sharp Fermi-gas boundary,
about $30\%$ of the events have higher recoil energy than the LZ event.
Halo DM scattering would add little extra energy resulting in a similar spectrum, unless the interaction significantly grows with momentum.

%\footnote{Unlike in~\cite{2112.09111} we here assume that such processes are easily kinematically open, even in nuclei.}
%\beq
%n \to \chi \chi\chi ~\cite{2112.09111},\qquad
%\bar\chi n\longrightarrow\chi \chi ~\cite{2112.09111},\qquad
%\chi n\to \chi\nu~\cite{2406.00445}.
%\eeq

\begin{table}
\begin{center}
\begin{tabular}{@{}llllll@{}}
%\toprule
Experiment & Mode & Limit in yr & Target & Signature & Reference\\
\midrule
SNO+ (2022) & $n\to\text{inv}$ & $9.0 ~ 10^{29}$ & $^{16}$O & De-excitation $\gamma$ & \cite{2205.06400}\\
%SNO+ (2022) & $p\to\text{inv}$ & $9.6 ~ 10^{29}$ & $^{16}$O & De-excitation $\gamma$ & \cite{2205.06400}\\
KamLAND (2006) & $n\to\text{inv}$ & $5.8~ 10^{29}$ & $^{12}$C & Correlated de-excitation & \cite{KamLAND:2005pen}\\
%DAMA/LXe (2000) & $p\to\text{inv}$ & $1.9 ~10^{24}$ & $^{129}$Xe & Delayed daughter decay & \cite{DAMA}\\
%JUNO (proj.) & $n\to\text{inv}$ & $5.0 ~ 10^{31}$ & $^{12}$C & Triple coincidence & \cite{JUNO}\\
%\midrule
{LZ (2026)} & $n\to\text{inv}$ & ${2.0 ~ 10^{29}}$ & $^{\rm nat}$Xe & {Prompt nuclear recoil} & This work\\
%\bottomrule
\end{tabular}
\caption{\em\label{tab:inv} Limits at $90\%$ confidence level on neutron disappearance in different nuclei.}
\end{center}
\end{table}

\subsection*{Bounds on invisible neutron decay}
The neutron disappearance interpretation of the LZ anomaly 
needs an effective  lifetime $\tau_n \approx 8~10^{29}\,{\rm yr}$ for neutrons inside nuclei.
This  is viable, altought constrained.
Indeed, the SNO~\cite{2205.06400} and KamLAND~\cite{KamLAND:2005pen} limits on invisible neutron disappearance in table~\ref{tab:inv}
permit about $1.1$ neutron disappearances in the LZ exposure, $ 1.0~ 10^{30}\ \text{neutron yr}$.
The same situation arises for ${\rm DM}\, n$ induced neutron disappearance~\cite{2112.09111},
after recasting $\tau_n$ into the cross section $\sigma_{{\rm DM}n} = 2 M_{\rm DM} /\tau_n \rho_\odot \langle v\rangle \approx (6~ 10^{-45}\,{\rm cm}^2)(M_{\rm DM}/\GeV)$
with DM density $\rho_{\rm DM}\approx 0.4\GeV/{\rm cm}^3$ and velocity $\langle v\rangle\approx 10^{-3}$.

\smallskip

At fundamental level, neutron disappearance arises from operators annihilating three quarks, 
that generically also generate visible processes, such as $n\to\pi^0\slashed E$ and $p\to\pi^+ \slashed E$.
Visible final states are typically excluded by Super-Kamiokande~\cite{1508.05530}, which has $10^5$ larger exposure than LZ.
Avoiding the Super-Kamiokande exclusion needs a mass range such that these visible processes are kinematically closed: 
the energy released must be  less than a pion mass.
This implies process-dependent mass ranges that comfortably allow the release of the $E_R$ claimed by LZ.
% for the decay:
%Since the observed recoil demands $M_{\rm inv}\lesssim898\MeV$ while closure of $p\to\pi^++\slashed E$ on free protons
%requires $M_{\rm inv}>m_p-m_{\pi^+}=798.7\MeV$, the invisible final state must satisfy
%\beq
%799\MeV\lesssim \sum m_{\rm inv}\lesssim 898\MeV.
%\eeq
%Neutron stars are unaffected.

\subsection*{Extra nuclear activity}
A possible difficulty with the neutron-disappearance interpretation is the additional signal produced by nuclear de-excitation. 
Removing a bound neutron generally leaves a neutron-hole state, which can decay by emitting $\gamma$ rays, internal-conversion electrons, or additional nucleons. 
Depending on the hole state, the released energy can range from tens of keV to several MeV~\cite{2112.09111,2406.00445}. 
Such a prompt decay would negligibly shift the nucleus momentum giving an event that likely does not appear in LZ as an isolated nuclear recoil.
Indeed LZ measures both prompt scintillation and ionization and classified the event as a single-site, 
nuclear-recoil-like interaction with no coincident signal above the veto thresholds~\cite{2609.02823}.
This constrains prompt accompanying radiation, although LZ did not explicitly test mixed nuclear-recoil plus nuclear-de-excitation signals.
A small de-excitation deposit might remain unresolved, biasing the reconstructed recoil energy.

\begin{table}[t]
\centering
\small
\begin{tabular}{@{}cccccc|c@{}}
\toprule
Parent & Abundance & Daughter state & Level & Excitation
& Half-life & $E_R$ \\
\midrule
$^{132}\mathrm{Xe}(0^+)$
& $27\%$
& $^{131}\mathrm{Xe}(11/2^-)$
& isomer
& $164\keV$ & 11.8 day & $230\keV$ \\
&
&
$^{131}\mathrm{Xe}(3/2^+)$
& ground state
& $~~0\keV$ & stable & $237\keV$ \\[1mm]

$^{134}\mathrm{Xe}(0^+)$
& $10\%$
& $^{133}\mathrm{Xe}(11/2^-)$
& isomer
& $233\keV$ & 2.20 day & $241\keV$ \\
&
&
$^{133}\mathrm{Xe}(3/2^+)$
& ground state
& $~~0\keV$ & 5.25 day & $252\keV$ \\[1mm]

$^{136}\mathrm{Xe}(0^+)$
& $9\%$
& $^{135}\mathrm{Xe}(11/2^-)$
& isomer
& $527\keV$ & 15 min & $245\keV$ \\
&
&
$^{135}\mathrm{Xe}(3/2^+)$
& ground state
& $~~0\keV$ & 9.14 hr & $269\keV$ \\[1mm]

$^{130}\mathrm{Xe}(0^+)$
& $4\%$
& $^{129}\mathrm{Xe}(11/2^-)$
& isomer
& $236\keV$ & 8.9 day & $214\keV$ \\
&
&
$^{129}\mathrm{Xe}(1/2^+)$
& ground state
& $~~0\keV$ & stable & $226\keV$ \\
\bottomrule
\end{tabular}
\caption{\label{tab:xe}\em Quiet daughter states of the dominant even-$A$ xenon
isotopes. The recoil energies in the last column assume a special near-threshold model with $m_\phi=6\GeV$, $Q=14\MeV$,
and negligible incident-particle momentum.}

\end{table}

Assuming that nuclear activity is incompatible with the LZ observation,
we estimate a $10\%$ probability that neutron disappearance left a long-lived xenon isomer that avoids a prompt decay. 
Indeed, natural xenon is a mixture of various isotopes.
Those with odd $A$ form $50\%$ of the total abundance and typically give a prompt $\gamma$;
the probability of leaving the daughter in its ground state is at $\%$ level.
The other half is formed by even-$A$ xenon isotopes in $0^+$ ground states: 
if a neutron with large angular momentum $j=\ell +1/2$ is removed, the resulting odd-$A$ daughter 
can be left in a state with total spin $J=j$ that undergoes a slow decay.
In particular, removing the neutron from the $1h_{J}$ orbital (we number radial orbitals starting from one)
with  highest $J = 11/2$ can lead to decays necessarily suppressed by high electromagnetic multipolarity $(E_\gamma R)^{2(\ell-1)}$.
The dominant Xe isotopes have the comfortably long half-lives~\cite{nuclear} listed in table~\ref{tab:xe}.
%\begin{center}
%\begin{tabular}{ccccc}
%%\toprule
%parent & abundance &  isomer & excitation $E_f^*$ & half-life\\
%\midrule
%$^{132}\mathrm{Xe}(0^+)$ & 27\% &  $^{131m}\mathrm{Xe}(11/2^-)$ & $164\keV$ & $11.8$ d\\
%$^{134}\mathrm{Xe}(0^+)$ & 10\% &  $^{133m}\mathrm{Xe}(11/2^-)$ & $233\keV$ & $2.20$ d\\
%$^{136}\mathrm{Xe}(0^+)$ & 9\% &  $^{135m}\mathrm{Xe}(11/2^-)$ & $527\keV$ & $15$ min\\
%$^{130}\mathrm{Xe}(0^+)$ & 4\% & $^{129m}\mathrm{Xe}(11/2^-)$ & $236\keV$ & $8.9$ d\\
%%\bottomrule
%\end{tabular}
%\end{center}
Furthermore, removing a $2d_{3/2}$ neutron from $^{132,134,136}\mathrm{Xe}$ can leave a $^{131,133,135}\mathrm{Xe}$ in its $3/2^+$ ground state.
Removing a $3s_{1/2}$ neutron from $^{130}\mathrm{Xe}$ can leave a $^{129}\mathrm{Xe}$ in its $1/2^+$ ground state.
A naive counting of the occupancy fraction of the neutrons leads to the $10\%$ estimate for the invisible fraction.
This fraction grows to $\approx 14\%$ by restricting to the LZ energy, $p\approx 250 \MeV$.
More precise computations might find order one corrections.
A related isomer-production process is experimentally established~\cite{Luo2022}. 
Taking into account the estimated quiet fraction allows $N_{\rm quiet}\lesssim 0.1$ neutron disappearance with no prompt extra signal,
namely a 1 in 10 coincidence.

\subsection*{Near-threshold induced neutron decay}
This mildly problematic bound gets relaxed or avoided if neutron disappearance is induced and provides limited energy.
Indeed SNO and KamLAND search for neutron disappearance in oxygen and carbon, where removing a
neutron costs $S_n(^{16}{\rm O})=15.7\MeV$ and $S_n(^{12}{\rm C})=18.7\MeV$, about twice as much as in xenon where
$S_n({\rm Xe})\approx 9\MeV$.
If a process ${\rm DM} \, n \to \hbox{invisible}$ releases between  $9\MeV$ and  $15.6\MeV$,
it is kinematically open in xenon (and in argon and germanium) and closed in oxygen and carbon.
The same restriction suppresses xenon channels involving deep neutrons that produce a daughter nucleus with large excitation energy,
favouring valence orbitals and final low-lying states.
The quiet fraction increases up to $\sim30\%$ and  the recoil spectrum of fig.\fig{LZn} gets significantly modified.
For example, introducing an extra fermion $\chi$ and scalar $\phi$,  
a $\chi n \to\phi^*$ operator $\phi\chi \,udd /\Lambda^3$  with 
$m_\chi\approx 5\GeV$, $ Q\equiv m_\chi + m_n -m_\phi \approx 14\MeV$, $\Lambda \approx 500\TeV$ (a value allowed by collider constraints)
would give co-stable invisible $\chi$ and $\phi$; 
the two body  kinematics turns the $E_R$ spectrum into narrow nuclear lines, 
with predicted intensity and positions as function of the masses, as we now compute.
Energy and momentum conservation in the $\chi~ {}^A {\rm Xe} \to \phi^\ast ~ {}^{A-1}{\rm Xe}_f$ reaction read
\begin{equation}
    Q  = S_n + E_\phi + E_R   + E_f^\ast,\qquad
    0=\vec{p}_{A-1}+\vec{p}_\phi.
\end{equation}
We neglected the small kinetic energy of the incoming particle, so that
the energy $Q$ gained by converting a free neutron is partially used 
for the neutron separation energy $S_n$, 
for the recoil kinetic energy of the invisible particle $E_\phi$,
of the daughter nucleus $E_R$,
and for its excitation energy $E_f^\ast$.
In the non-relativistic limit momentum conservation allows to eliminate $E_\phi =E_R M_{A-1}/m_\phi$ obtaining
\begin{equation}
    E_{R} \approx \frac{Q-S_n-E_f^\ast}{1+M_{A-1}/m_\phi}.
\end{equation}
This shows that the reaction is kinematically allowed if $Q > S_n + E_f^\ast$.
Demanding closure in SNO limits $Q< S_n(^{16}{\rm O})$ so that getting
the LZ recoil energy $E_R$ needs $m_\phi\gtrsim5\GeV$.
This mass range and the involvement of baryon number would point to asymmetric DM models.
Assuming $m_{\phi}=6\GeV$ and $Q=14\MeV$, so that $m_\chi\approx 5\GeV$, 
we list in table~\ref{tab:xe} the main quiet channels,
ground state and long-lived isomers, showing that they give lines near the observed $E_R\approx 248 \keV$.
The prompt channels too are illustrated in the right panel of fig.\fig{LZn}.
No channel emits a neutron and all non-quiet events consist of
a nuclear recoil accompanied by a $\gamma$ cascade. 
%Correlated pairs, which would require $\gtrsim60\MeV$, do not contribute. 
%Fragmentation of the hole strength, with spectroscopic factors $\sim0.6$, broadens the deeply-bound lines and reduces the quiet fraction to $\sim15-25\%$.
Including the small initial DM momentum leads to several keV broadening and to annual variation in the line positions
(rather than in the intensities).

\subsection*{Conclusions and predictions}
The neutron-disappearance interpretation of the LZ event can be readily tested.
Within LZ itself the energy spectrum is predicted to be as
in the left panel of fig.\fig{LZn} when the final-state phase space is unrestricted.
In this case most nuclear recoils should be accompanied by prompt nuclear de-excitation, typically including MeV $\gamma$. 
About $14\%$ of the events near the observed recoil energy are instead expected to populate long-lived isomers and therefore remain prompt-quiet; their subsequent decays produce delayed $\gamma$ rays or conversion electrons.
The neutron Fermi momenta are comparable in different nuclei, 
whereas the recoil energy scales approximately as $E_R\approx p^2/2M_{A-1}$,
giving endpoints around
\beq
E_R^{\rm max}\approx 0.21,\ 0.29,\ 0.51,\ 0.92,\ 1.25\MeV
\quad\hbox{in}\qquad\hbox{ W, Xe, Ge, Ar, Si}.
\eeq
So DarkSide, SuperCDMS and CRESST probe the same physics at energies above their
usual windows. 
This interpretation needs a neutron disappearance rate around SNO limits, that can be
significantly improved by JUNO and THEIA~\cite{2405.17792}.
A restricted kinematics avoids SNO and KamLAND bounds on neutron disappearance,
increases the fraction of quiet events, modifies the $E_R$ spectrum
as exemplified in the right panel of fig.\fig{LZn}.
Neutron decay predicts no annual modulation and no dependence on the halo, 
while induced neutron disappearance predicts the standard few-\% modulation.

\medskip

If the LZ anomaly is not confirmed, this analysis sets the limit in table~\ref{tab:inv} on  neutron disappearance.
Despite being weakened by one event,
this is the leading bound on near-threshold, 
DM-induced neutron disappearance into invisible final states,
not tested by experiments using oxygen and carbon such as SNO and KamLAND.

\subsection*{Acknowledgement}
M.~Aghaie is supported by JSPS KAKENHI grant numbers 24H02244 and 26K24541.
We thank ChatGPT and Claude for useful discussions.

\end{document}